\documentclass[12pt]{article}
\catcode`\@=11
\@addtoreset{equation}{section}

\global\arraycolsep=2pt
\usepackage{mathrsfs,amsbsy,amssymb,latexsym,amsfonts,amsmath,cite}
\usepackage{graphicx,color}
\usepackage{physics}
\usepackage{mathtools}
\usepackage{caption}
\usepackage{here,comment}

\usepackage{mathrsfs,amsbsy,amssymb,latexsym,amsfonts,amsmath,cite,bm}
\usepackage{graphicx,color}
\usepackage{mathtools}
\usepackage{enumerate}

\DeclareFontFamily{U}{BOONDOX-calo}{\skewchar\font=45 }
\DeclareFontShape{U}{BOONDOX-calo}{m}{n}{
  <-> s*[1.05] BOONDOX-r-calo}{}
\DeclareFontShape{U}{BOONDOX-calo}{b}{n}{
  <-> s*[1.05] BOONDOX-b-calo}{}
\DeclareMathAlphabet{\mathcalboondox}{U}{BOONDOX-calo}{m}{n}
\SetMathAlphabet{\mathcalboondox}{bold}{U}{BOONDOX-calo}{b}{n}
\DeclareMathAlphabet{\mathbcalboondox}{U}{BOONDOX-calo}{b}{n}

\newcommand{\no}{\nonumber}

\newcommand{\cF}{\mathcal F}

\newcommand{\cL}{\mathcal L}
\newcommand{\cM}{\mathcal M}

\newcommand{\cv}{\mathcalboondox v}

\newcommand{\pa}{\partial}

\allowdisplaybreaks

\begin{document}

\renewcommand{\thefootnote}{\fnsymbol{footnote}}

\begin{flushright} 
RIKEN-iTHEMS-Report-24, STUPP-24-270
\end{flushright}
\vspace*{0.5cm}

\begin{center}
{\Large \bf  4D Chern-Simons theory with auxiliary fields
}
\vspace*{2cm} \\
{\large  Osamu Fukushima$^{\sharp}$\footnote{E-mail:~osamu.fukushima@riken.jp},
and Kentaroh Yoshida$^{\natural}$\footnote{E-mail:~kenyoshida@mail.saitama-u.ac.jp
}} 
\end{center}

\vspace*{0.4cm}

\begin{center}
$^{\sharp}${\it iTHEMS, RIKEN, Wako, Saitama 351-0198, Japan}
\end{center}
\begin{center}
$^{\natural}${\it Graduate School of Science and Engineering, Saitama University, 255 Shimo-Okubo, Sakura-ku, Saitama 338-8570, Japan}
\end{center}

\vspace{2cm}

\begin{abstract}
The auxiliary field sigma model (AFSM) has recently been constructed by Ferko and Smith as deformations of the principal chiral model by including auxiliary fields and the potential term given by an arbitrary univariate function. This AFSM provides an infinite family of integrable sigma models including the original $T\overline{T}$-deformation and the root $T\overline{T}$-deformation. In this paper, we propose a 4D Chern-Simons (CS) theory with auxiliary fields. Then the AFSM is derived from this CS theory with the twist function for the principal chiral model by imposing appropriate boundary conditions for the gauge field and auxiliary fields. We also derive the AFSM with the Wess-Zumino term by deforming the twist function and modifying the boundary conditions.

\end{abstract}

\setcounter{footnote}{0}
\setcounter{page}{0}
\thispagestyle{empty}

\newpage

\tableofcontents

\renewcommand\thefootnote{\arabic{footnote}}

\section{Introduction}\label{sec:introduction}

A fascinating topic in mathematical physics is to study integrable models. Although it has a long history, new integrable models are still being discovered today. In this paper, we will focus upon two issues: 1) 4D Cherns-Simons (CS) theory and 2) auxiliary field sigma model (AFSM). 

\medskip

The first one, 4D CS theory was proposed by Costello and Yamazaki \cite{Costello:2019tri} as a candidate of the unified theory of 2D integrable models\footnote{For another scenario based on the affine Gaudin model, see a series of papers \cite{Vicedo:2017cge,Vicedo:2019dej,Lacroix:2020flf}.}. One can reproduce integrable models such as 2D principal chiral models (with the Wess-Zumino term) \cite{Delduc:2019whp}, symmetric coset sigma models \cite{Fukushima:2020dcp} and integrable deformations of them 
like Yang-Baxter deformations\footnote{For the original works on Yang-Baxter deformations, see \cite{Klimcik:2002zj,Klimcik:2008eq}. For a pedagogical book, see \cite{Yoshida}.}  \cite{Delduc:2019whp,Fukushima:2020dcp} as well as the Faddeev-Reshetikhin model \cite{Fukushima:2020tqv}, non-abelian Toda field theories including (complex) sine-Gordon model and Liouville field theory \cite{Fukushima:2021ako}. For other related topics, see \cite{Schmidtt:2019otc,Tian:2020ryu,Tian:2020pub,Fukushima:2020kta,Caudrelier:2020xtn,Fukushima:2021eni,Stedman:2021wrw,Vicedo:2022mrm,Liniado:2023uoo,Boujakhrout:2023qlz,Levine:2023wvt,Ashwinkumar:2023zbu,Berkovits:2024reg,Schenkel:2024dcd,Chen:2024axr}. 
For a concise review, see \cite{Lacroix:2021iit}. 





\medskip 

The second one, AFSM has recently been presented by Ferko and Smith \cite{Ferko:2024ali}\footnote{For earlier works along the similar direction, see \cite{Ivanov:2001ec,Ivanov:2002ab,Ivanov:2003uj,Ivanov:2012bq}}. 
This is a generalization of the principal chiral model by including auxiliary fields and an arbitrary interaction function. This AFSM provides an infinite family of new integrable sigma models. More interestingly, intriguing integrable deformations like $T\overline{T}$-deformations \cite{Smirnov:2016lqw, Cavaglia:2016oda} and root $T\overline{T}$-deformations \cite{Ferko:2022cix,Babaei-Aghbolagh:2022leo,Babaei-Aghbolagh:2022uij} are included in the AFSM, as noted in \cite{Ferko:2024ali}.   

\medskip 

The aim of this paper is to derive the AFSM from a 4D CS theory with auxiliary fields. We will generalize the original 4D CS theory by including auxiliary fields and the potential term given by an arbitrary univariate smooth function. 
We will refer to this model as the 4D auxiliary field Chern-Simons theory (AFCST).
Then we will derive the AFSM from the AFCST by imposing appropriate boundary conditions for the gauge field and auxiliary fields. The twist function is the same as that of the principal chiral model. It is also possible to include the Wess-Zumino term in this analysis by deforming the twist function as in \cite{Delduc:2019whp}. As a result, we can derive the AFSM with the Wess-Zumino term. 

\medskip 

This paper is organized as follows. In section 2, we shall give a brief review of the AFSM. In section 3, we will present the AFCST. Then the AFSM will be derived from the AFCST by using the twist function for the principal chiral model. We also generalize this analysis by including the Wess-Zumino term. 
Section 4 is devoted to conclusion and discussion.

\section{An infinite family of integrable sigma models}\label{sec:infinite}

Before delving into the derivation of the auxiliary field sigma model (AFSM)~\cite{Ferko:2024ali} from the 4D CS theory, 
let us briefly review the AFSM itself, which can be regarded as an infinite family of integrable deformations of the principal chiral model (PCM).
As we will discuss later, the flatness condition of the Lax pair for the AFSM reproduce its equations of motion only under the constraints by auxiliary equations, in contrast to integrable systems in the standard sense.

\medskip

The AFSM is defined on 2D Minkowski spacetime $\cM$ with coordinates $\sigma^{\pm}$\,.
The physical degrees of freedom are provided through a group-valued field $g:\cM\to G$\,, where $G$ is a Lie group associated with a Lie algebra $\mathfrak{g}$\,.

\medskip

The classical action of the AFSM is given by 
\begin{align}
    S_{\rm AFSM}[g,v_{\pm}]:=&\,
    \int_{\cM}d\sigma^+\wedge d\sigma^- 
    \bigg(
    \frac{1}{2}\tr(j_+j_-) - \tr(v_+v_-) + \tr(j_+v_- - j_-v_+) + E(\nu)
    \bigg)\,,
    \label{auxiliary-2d}
    \\
    j_{\pm}:=&\,
    g^{-1}\pa_{\pm}g\,,
    \qquad
    \nu:=\tr(v_+v_+)\tr(v_-v_-)\,,
    \label{action-AFSM}
\end{align}
where $E(\nu)$ is an arbitrary univariate smooth function.
The auxiliary fields $v_{\pm}:\cM\to\mathfrak{g}$ are defined as Lie algebra valued fields.
The components of the Maurer-Cartan one-form, $j_{\pm}$ satisfy the off-shell flatness condition
\begin{align}
    0 = \pa_{+}j_{-}-\pa_{-}j_{+} + [j_{+},j_{-}] \,.
    \label{off-shell-flatness}
\end{align}

\paragraph{Equations of motion}

Let us next consider the equations of motion for $S_{\rm AFSM}$\,.  
The variation for the auxiliary fields $v_{\pm}\mapsto v_{\pm}+\delta v_{\pm}$
leads to
\begin{align}
    0=&\, \pm j_{\pm} - v_{\pm} + 2 v_{\mp}\tr(v_{\pm}v_{\pm})E'(\nu)\,,
    \label{auxiliary-eom}
\end{align}
where $E'(\nu)$ denotes the derivative of $E(\nu)$\,. 
Then the variation $g\mapsto g+\delta g=g + \epsilon g$ 
gives rise to the equation of motion
\begin{align}
    0=&\, \pa_{+}j_{-} + \pa_{-}j_{+}
    -2\big(
    [v_{-},j_{+}] - [v_{+},j_{-}] - \pa_{+}v_{-} + \pa_{-}v_{+}
    \big)\,.
    \label{2d-eom}
\end{align}
It is helpful to define the symbol $\dot{=}$ as the equality that holds under the auxiliary equation (\ref{auxiliary-eom}).
From the relation (\ref{auxiliary-eom}), one can see the following relations: 
\begin{align}
    [v_-,j_+] \,\dot{=}\, [v_{+},j_-]  \,\dot{=}\, -[v_+,v_-] \,.
    \label{auxiliary-com}
\end{align}
By taking account of the above relations, the equation of motion (\ref{2d-eom}) can be rewritten into a local conservation form:
\begin{align}
    0=&\,
    \pa_{+}\mathfrak{J}_{-} + \pa_{-}\mathfrak{J}_{+}\\
    =&\,
    \pa_{+}\big(-j_{-} - 2v_{-}\big) + \pa_{-}\big(-j_{+} + 2v_{+}\big)\,,
\end{align}
where the modified current $\mathfrak{J}$ is defined as 
\begin{align}
    \mathfrak{J}_{\pm}:= &\,
    - \big(j_{\pm} \mp 2 v_{\pm}\big)\,. 
    \label{modified-def}
\end{align}

\paragraph{Trivial case}

Note here that AFSM with the trivial potential $E'(\nu)=0$ is equivalent to the standard principal chiral model 
\[
S_{\rm PCM}[g]:=-\int_{\cM}d\sigma^+\wedge d\sigma^{-}
\frac{1}{2}\tr(j_{+}j_{-})\,.
\]
This is because the auxiliary equations mean $j_{\pm}\dot{=}\pm v_{\pm}$ when $E'(\nu)=0$\,. Hence the classical action is rewritten as  
\begin{align}
    S_{\rm AFSM}[g, \pm j_{\pm}] =&\,
    \int_{\cM}d\sigma^{+} \wedge d\sigma^{-}
    \left(
    \frac{1}{2}\tr(j_{+}j_{-}) + \tr(j_{+}j_{-})
    -2 \tr(j_{+}j_{-}) + E(\nu)
    \right)\no\\
    =&\,
    S_{\rm PCM}[g] + \mathrm{const.}\,. 
\end{align}

\paragraph{Lax pair}

The Lax pair for the AFSM (\ref{auxiliary-2d}) is given by
\begin{align}\begin{split}
    \mathscr{L}_{\pm}:=&\,
    \frac{j_{\pm}\pm z \mathfrak{J}_{\pm}}{1-z^2}\\
    =&\,
    \frac{1}{1\pm z}j_{\pm} + \frac{2z}{1-z^2}v_{\pm}\,,
    \label{auxiliary-Lax}
\end{split}\end{align}
with the spectral parameter $z\in\mathbb{C}$\,.
It satisfies the on-shell flatness condition
\begin{align}
    0\,\dot{=}\,&\,\pa_{+}\mathscr{L}_{-} - \pa_{-}\mathscr{L}_{+} + [\mathscr{L}_{+},\mathscr{L}_-]\,.
    \label{AFSM-flatness}
\end{align}
Note that the equality here is given by $\dot{=}$ rather than $=$ because the right-hand side of (\ref{AFSM-flatness}) is evaluated as 
\begin{align}
    &\pa_{+}\mathscr{L}_{-} - \pa_{-}\mathscr{L}_{+} + [\mathscr{L}_{+},\mathscr{L}_-]
    \no\\
    &=
    \frac{\pa_{+}j_{-}-\pa_{-}j_{+} -z (\pa_{+}\mathfrak{J}_{-} + \pa_{-}\mathfrak{J}_{+})}{1-z^2}
    +
    \frac{[j_+,j_-] - z[j_{+},\mathfrak{J}_-]
    + z[\mathfrak{J}_{+},j_{-}] -z^2 [\mathfrak{J}_+,\mathfrak{J}_-]}
    {(1-z^2)^2}
    \no\\
    &\,\dot{=}\,
    \frac{\pa_{+}j_{-}-\pa_{-}j_{+} -z (\pa_{+}\mathfrak{J}_{-} + \pa_{-}\mathfrak{J}_{+})}{1-z^2}
    +\frac{[j_+,j_-]}{1-z^2}
    \no\\
    &=
    \frac{-z}{1-z^2}\big( \pa_{+}\mathfrak{J}_{-} + \pa_{-}\mathfrak{J}_{+} \big)\,.
\end{align}
In the third line, we have utilized the commutation relations
\begin{align}
    [j_{+},\mathfrak{J}_-] \,\dot{=}\, [\mathfrak{J}_{+},j_-]\,,
    \qquad
    [j_{+},j_{-}] \,\dot{=}\, [\mathfrak{J}_{+},\mathfrak{J}_{-}]\,,
    \label{jj-relation}
\end{align}
which follows from (\ref{auxiliary-com}) and (\ref{modified-def}).

\medskip

We stress that the flatness condition for the Lax pair (\ref{AFSM-flatness}) only gives parts of the equations of motion, and does not reproduce any information about the auxiliary equations (\ref{auxiliary-eom}).
Rather, the equivalence of the equations of motion and (\ref{AFSM-flatness}) would hold in principle after deleting the auxiliary fields $v_{\pm}$ by substituting the solution of (\ref{auxiliary-eom}).
Thus, the AFSM (\ref{action-AFSM}) is integrable in this sense.

\section{AFSM from AFCST}\label{sec:auxiliary-4dCS}

In this section, we introduce the action of the AFCST.
Appropriately solving a part of the equations of motion, we can reduce the 4D action into the 2D action (\ref{auxiliary-2d}) accompanied with the Lax pair (\ref{auxiliary-Lax}).

\subsection{The classical action of the AFCST}

The 4D CS theory is defined on the four-dimensional space $\cM\times C$\,, where $C=\mathbb{C}P^1$ is the complex projective space with the complex coordinates $(z,\bar{z})$\,. Usually, the CS theory is defined in odd dimensions by using the CS form. Hence we need to use an additional one-form so as to define the classical action of the 4D CS theory. In the following, the one-form is taken as a meromorphic one-form 
\begin{align}
    \omega:=&\, \varphi(z)dz\,,\qquad
    \varphi(z) = \frac{1-z^2}{z^2}\,. 
    \label{twist}
\end{align}
Here $\varphi(z)$ is called the twist function which is closely related to the Poisson structure of the underlying integrable field theory~\cite{Vicedo:2017cge,Vicedo:2019dej}. It should be remarked that the twist function $\varphi(z)$ here is the same as that of the principal chiral model and the AFSM \cite{Ferko:2024ali}. As a matter of course, it may be more general from the viewpoint of the 4D CS theory. This issue will be discussed in another place \cite{FY}.

\subsubsection*{Classical action}

In order to derive the AFSM from a possible 4D CS theory, we would like to deform the original 4D CS theory by including auxiliary fields as in the AFSM. 

\medskip 

Our proposal for the AFCST is the following: 
\begin{align}
S_{\rm tot}[\cv,A] =&\, S_{\rm 4dCS}[A] + S_{\rm int}[\cv,A] + S_{\rm pot}[\cv]\,,
\label{S-tot}
\\
S_{\rm 4dCS}[A]:=&\,\frac{i}{4\pi}\int_{\cM\times C}\omega\wedge
\tr(A\wedge dA+ \frac{2}{3}A\wedge A\wedge A )\,,
\label{S-4dCS}
\\
S_{\rm int}[\cv,A]:=&\,
    \frac{i}{\pi}\int_{\cM\times C}\omega\wedge \tr( F(A)\wedge \cv)
    +\frac{i}{\pi}\int_{\cM\times C}\omega\wedge 
    \tr(\cv\wedge d\cv + 2 A\wedge \cv\wedge \cv)
    \,,
\label{S-int}
\\
S_{\rm pot}[\cv]:=&\,
2\int_{\cM\times C}dz\wedge d\bar{z} \wedge d\sigma^+ \wedge d\sigma^-
\sum_{\hat{z}\in\{\pm 1\}}\delta(z-\hat{z})
E(\xi)\no\\
=&\,
2\int_{\cM}d\sigma^+\wedge d\sigma^-\sum_{\hat{z}\in\{\pm 1\}}\lim_{z\to\hat{z}}E(\xi)
\label{S-pot}\\
\begin{split}
F(A):=&\,dA +  A\wedge A\,,
\\
\xi:=&\,
\frac{16(1-z^2)^2}{z^4}\,\tr(\overline{\cv_+}\cv_+)\tr(\overline{\cv_-}\cv_-)
=16\varphi^2(z)\tr(\overline{\cv_+}\cv_+)\tr(\overline{\cv_-}\cv_-)\,,
\end{split}\label{xi-def}
\end{align}
Here $A:\cM\times (C\backslash\{\pm 1\})\to\mathfrak{g}^{\mathbb{C}}$ denotes the gauge field, and $\cv_{\pm}:\cM\times (C\backslash\{\pm 1\})\to \mathfrak{g}^{\mathbb{C}}$ are $\mathfrak{g}^{\mathbb{C}}$-valued fields. Note that the one-form $\cv=\cv_+d\sigma^++\cv_-d\sigma^-$ has no components in the $C$ direction.
%
%
The first term (\ref{S-4dCS}) is the original 4D CS action~\cite{Costello:2019tri} with the twist function (\ref{twist}).
The second term (\ref{S-int}) includes the interaction between $\cv$ and the gauge field $A$, and it respects the gauge invariance that will be discussed later. 

\medskip

The potential term (\ref{S-pot}) requires slight caution because the argument of the arbitrary function $E$ incorporates a $(1-z^{2})^2$ factor, while the action includes the delta function $\delta(z\pm1)$.
Since the $\mathfrak{g}^\mathbb{C}$-valued field $\cv$ is defined on $\cM\times (C\backslash\{\pm 1\})$\,, it allows singularities $z=\pm 1$\,.
Although the argument $\xi$ appears to tend to zero in the $z\to \pm 1$ limit, $\xi$ gives nontrivial contributions because of this singularities. The overline denotes the complex conjugation.

\subsubsection*{Reality conditions}
Although the gauge field $A$ and the one-form field $\cv$ take values in the complexified Lie algebra $\mathfrak{g}^\mathbb{C}$, it is necessary to ensure the reality of the actions (\ref{S-4dCS}), (\ref{S-int}), and (\ref{S-pot}) by imposing an appropriate reality condition on the fields.
We describe the reality condition by introducing an involutive anti-linear automorphism $\mu_{\rm t}:C\to C$
\begin{align}
    \mu_{\rm t}:z \mapsto \bar{z}\,,
\end{align}
and an involutive automorphism $\tau:\mathfrak{g}^{\mathbb{C}}\to \mathfrak{g}^{\mathbb{C}}$ such that
\begin{align}
    \overline{\tr(\mathsf{x}\,\mathsf{y})} = \tr(\tau(\mathsf{x})\,\tau(\mathsf{y}))\,,\qquad
    \mathsf{x},\mathsf{x}\in\mathfrak{g}^{\mathbb{C}}\,.
\end{align}
If the following equivariance conditions holds:
\begin{align}
    \mu_{\rm t}^* A = \tau(A)\,, \qquad
    \mu_{\rm t}^* \cv = \tau(\cv)\,, 
    \label{reality}
\end{align}
all terms of the action are shown to be real, i.e.,
\begin{align}
    \overline{S_{\rm 4dCS}[A]} = S_{\rm 4dCS}[A]\,, \qquad
    \overline{S_{\rm int}[\cv,A]} = S_{\rm 4dCS}[\cv, A]\,, \qquad
    \overline{S_{\rm pot}[\cv]} = S_{\rm pot}[\cv]\,.
\end{align}
One can confirm this statement by utilizing facts such as the twist function (\ref{twist}) satisfying $\overline{\omega}=\mu_{\rm t}^{*}\omega$\,, and that the involution $\mu_{\rm t}$ reverses the orientation of $C$ as
\begin{align}
    \overline{i\int_{C}dz\wedge d\bar{z} \;\cF(A,\cv) }=&\,
    -i \int_{C}dz\wedge d\bar{z} \,\cF(\tau(A),\tau(\cv))
    =- i \int_{C}dz\wedge d\bar{z} \;\mu_{\rm t}^*\cF
    \no\\
    =& -i \int_{\mu_{\rm t}C}dz\wedge d\bar{z} \;\cF
    =i \int_{C}dz\wedge d\bar{z} \;\cF
    \,,
\end{align}
where $\cF$ is an arbitrary function on $C$\,.
Note also that $A\,,\cv\in\mathfrak{g}$ for $z\in\mathbb{R}$ since the real Lie algebra $\mathfrak{g}$ is defined as the fixed points of $\tau$\,.

\subsubsection*{Equations of motion}

By varying the gauge field $A$ as $A\mapsto A +\epsilon$ and setting $\delta S_{\rm tot}[\cv,A]=0$, we obtain the  equations of motion at $z\in\mathbb{C}P^1\backslash\{0,\infty\}$:
\begin{align}
    0=&\,
    \varphi(z) \Big[ F(A) 
    +
    2(d\cv + A\wedge \cv + \cv\wedge A )
    +
    4 \cv\wedge \cv\Big]
    \label{bulk-eom-form}
    \,,
\end{align}
which we refer to as the bulk equations of motion.
The equation $\delta S_{\rm tot}[\cv,A]=0$ also yields surface terms at $z\in\{0,\infty\}$, which is the so-called boundary equation of motion\footnote{In the following discussion, we repeatedly use the formula
\begin{align}
    \pa_{\bar{z}}\frac{1}{z}=-2\pi i\delta(z)\,.
\end{align}
}
\begin{align}
     \begin{split}
    0=&\,
    \operatorname{res}_{z=0}\left(\frac{1-z^2}{z^2}\right)
    \epsilon^{ij}\tr(A_i\delta A_j)\big|_{z=0}
    +\operatorname{res}_{z=0}\left(z\frac{1-z^2}{z^2}\right)
    \epsilon^{ij}\pa_{z}\tr(A_i\delta A_j)\big|_{z=0}
    \\
    &+
    \operatorname{res}_{w=0}\left(\frac{1-w^2}{w^2}\right)
    \epsilon^{ij}\tr(A_i\delta A_j)\big|_{w=0}
    +\operatorname{res}_{w=0}\left(w\frac{1-w^2}{w^2}\right)
    \epsilon^{ij}\pa_{w}\tr(A_i\delta A_j)\big|_{w=0}
    \,,
    \\
    &+
    4\operatorname{res}_{z=0}\left(\frac{1-z^2}{z^2}\right)
    \epsilon^{ij}\tr(\cv_i\delta A_j)\big|_{z=0}
    +4\operatorname{res}_{z=0}\left(z\frac{1-z^2}{z^2}\right)
    \epsilon^{ij}\pa_{z}\tr(\cv_i\delta A_j)\big|_{z=0}
    \\
    &+
    4\operatorname{res}_{w=0}\left(\frac{1-w^2}{w^2}\right)
    \epsilon^{ij}\tr(\cv_i\delta A_j)\big|_{w=0}
    +4\operatorname{res}_{w=0}\left(w\frac{1-w^2}{w^2}\right)
    \epsilon^{ij}\pa_{w}\tr(\cv_i\delta A_j)\big|_{w=0}
    \\
    =&\,
    \epsilon^{ij}\pa_{z}\tr(A_i\delta A_j)\big|_{z=0}
    +
    \epsilon^{ij}\pa_{w}\tr(A_i\delta A_j)\big|_{w=0}
    +
    4\epsilon^{ij}\pa_{z}\tr(\cv_i\delta A_j)\big|_{z=0}
    +
    4\epsilon^{ij}\pa_{w}\tr(\cv_i\delta A_j)\big|_{w=0}\,,
    \end{split}
\end{align}
where $w:=1/z$ is a local coordinate around $z=\infty$\,.
We can satisfy the boundary equation of motion by choosing the Dirichlet boundary condition as
\begin{align}\begin{split}
    A_{\pm}\big|_{z=0,\infty}=&\,0\,,\qquad 
    \\
    \cv_{\pm}\big|_{z=0,\infty}=&\,0\,.
    \label{Dirichlet}
\end{split}\end{align}
The Dirichlet boundary condition of the gauge fields corresponds to the choice $(A_{\pm},\pa_{z}A_{\pm})\big|_{z=0,\infty}\in \{0\}\ltimes \mathfrak{g}_{\rm ab}:=\{ (0,x)\,|\,x\in\mathfrak{g}\} $ in the language of \cite{Delduc:2019whp}.

\medskip

The variations $\delta \cv_+$, $\delta \cv_-$ then lead to other equations of motion
\begin{align}
    0=&\,
    \delta \cv_{+}\varphi(z)\Big[ F(A)_{\bar{z}-} +
    2\big(\pa_{\bar{z}}\cv_{-} + [A_{\bar{z}},\cv_{-}]\big)\Big]
    +\!\!
    \sum_{\check{z}\in\{0,\infty\}}2\pi i\delta(z-\check{z})
    \operatorname{res}_{\check{z}}\left(z\frac{1-z^2}{z^2}\right)
    \pa_{z}\big(\delta \cv_{+}\cv_{-}\big)
    \no\\
    &-
    \sum_{\hat{z}\in\{\pm1\}}2\pi i\delta(z-\hat{z})
    \lim_{z\to\hat{z}}\Big[32\varphi^2(z)\delta \cv_{+}\cv_{+}\tr(\cv_-\cv_-)E'(\xi)
    \Big]
    \no\\
    \begin{split}
    =&\,
    \delta \cv_{+}\varphi(z)\Big[ F(A)_{\bar{z}-} +
    2\big(\pa_{\bar{z}}\cv_{-} + [A_{\bar{z}},\cv_{-}]\big)\Big]\\
    &-
    \sum_{\hat{z}\in\{\pm1\}}2\pi i\delta(z-\hat{z})
    \lim_{z\to\hat{z}}\Big[32\varphi^2(z)\delta \cv_{+}\cv_{+}\tr(\cv_-\cv_-)E'(\xi) \Big]
    \,,
    \end{split} \label{eom-v1}
    \\
    0=&\,
    \delta \cv_{-}\varphi(z) \Big[F(A)_{\bar{z}+} +
    2\big(\pa_{\bar{z}}\cv_{+} + [A_{\bar{z}},\cv_{+}]\big)
    \Big]
    +\!\!
    \sum_{\check{z}\in\{0,\infty\}}2\pi i\delta(z-\check{z})
    \operatorname{res}_{\check{z}}\left(z\frac{1-z^2}{z^2}\right)
    \pa_{z}\big(\delta \cv_{-}\cv_{+}\big)
    \no\\
    &-
    \sum_{\hat{z}\in\{\pm1\}}2\pi i\delta(z-\hat{z})
    \lim_{z\to\hat{z}}\Big[
    32\varphi^2(z)\delta \cv_{-}\cv_{-}\tr(\cv_+\cv_+)E'(\xi)
    \Big]
    \no\\
    \begin{split}
    =&\,
    \delta \cv_{-}\varphi(z)\Big[ F(A)_{\bar{z}+} +
    2\big(\pa_{\bar{z}}\cv_{+} + [A_{\bar{z}},\cv_{+}]\big)\Big]
    \,,\\
    &-
    \sum_{\hat{z}\in\{\pm1\}}2\pi i\delta(z-\hat{z})
    \lim_{z\to\hat{z}}\Big[
    32\varphi^2(z)\delta \cv_{-}\cv_{-}\tr(\cv_+\cv_+)E'(\xi)
    \Big]
    \,,
    \end{split}
    \label{eom-v2}
\end{align}
In deriving the equations (\ref{eom-v1}) and (\ref{eom-v2}), we used the boundary condition (\ref{Dirichlet}) on $\cv_{\pm}$\,.

\subsubsection*{Gauge invariance}
Upon imposing the boundary conditions (\ref{Dirichlet}), we now discuss the gauge invariance of the 4D CS theory.
Each term of the action (\ref{S-tot}) is indeed invariant under the following transformation
\begin{align}
    A\mapsto A^{u}:= uAu^{-1} - duu^{-1}\,,
    \qquad
    \cv \mapsto \cv^u:=u\cv u^{-1}\,,
    \label{standard-gauge}
\end{align}
where $u:\cM\times C\to G^{\mathbb{C}}$ is an arbitrary function subject to the constraint
\begin{align}
    u\big|_{z=0,\infty}= \bm{1}\,,
    \label{u-condition}
\end{align}
so that the boundary condition (\ref{Dirichlet}) holds.
In comparison with (\ref{standard-gauge}) and (\ref{u-condition}), transformations that do not satisfy the condition (\ref{u-condition}) are referred to as ``formal gauge transformations''.
Since the action is not invariant under a formal gauge transformation, it should rather be interpreted as a change of the variables into a new variable $u$\,.

\subsubsection*{Lax form}

By performing a formal gauge transformation, we define the Lax form by
\begin{align}
    A =:&\:
    \hat{g}\,\cL\,\hat{g}^{-1} - d\hat{g}\,\hat{g}^{-1}\,,
    \qquad
    \mbox{s.t. }\cL_{\bar{z}}=0\,.
    \label{Lax-form-def}
\end{align}
After performing this transformation, the one-form field $\cv$ is expressed as
\begin{align}
    \cv =:\hat{g}\,\hat{v}\,\hat{g}^{-1}\,.
\end{align}
Note here that the definition of the Lax form has an ambiguity:
\begin{align}
    \hat{g}\mapsto \hat{g}\,h\,,\qquad
    h:\cM\to G^{\mathbb{C}}\,,
    \label{2d-gauge}
\end{align}
which we call two-dimensional gauge transformations, distinguishing them from gauge transformations (\ref{standard-gauge}).
Under (\ref{2d-gauge}), the condition $\cL_{\bar{z}}=0$ is not altered since the following relation is satisfied 
\[
\pa_{\bar{z}}(\hat{g}h)h^{-1}\hat{g}^{-1}=\pa_{\bar{z}}\hat{g}\,\hat{g}^{-1}\,.
\]

\medskip

In terms of the Lax form (\ref{Lax-form-def}), the bulk equations of motion (\ref{bulk-eom-form}) read
\begin{align}
    0=&\, 
    \Big[\Big(\pa_{+}\cL_- -\pa_{-}\cL_+   + \big[\cL_+,\cL_-\big] \Big)
    + 2\big(\pa_{+}\hat{v}_- -\pa_{-}\hat{v}_+ 
    + [\cL_+,\hat{v}_-] + [\hat{v}_+,\cL_-] \big)
    + 4[\hat{v}_+,\hat{v}_-]\Big]\,,
    \label{Lax-flat-1}
    \\
    0=&\, 
    \varphi(z)\big[
    \pa_{\bar{z}}\cL_+  + 2\pa_{\bar{z}}\hat{v}_+ \big]
    \,,
    \label{Lax-flat-2}
    \\
    0=&\, 
    \varphi(z)\big[
    \pa_{\bar{z}}\cL_-  + 2\pa_{\bar{z}}\hat{v}_- \big]
    \,.
    \label{Lax-flat-3}
\end{align}
We kept the factor $\varphi(z)$ in (\ref{Lax-flat-2}) and (\ref{Lax-flat-3}) so that $\pa_{\bar{z}}\cL_\pm  + 2\pa_{\bar{z}}\hat{v}_\pm$ can have the delta function singularities at the zeros of $\varphi(z)$\,, i.e., $z=\pm1$\,.
We now introduce the \textit{modified Lax form} as
\begin{align}
    \mathfrak{L}_{\pm}:= \cL_{\pm} + 2\hat{v}_{\pm}\,,
\end{align}
and then the equations of motion (\ref{Lax-flat-1}) indicates the flatness condition:
\begin{align}
    0=&\, \pa_{+}\mathfrak{L}_{-} - \pa_{-}\mathfrak{L}_{+} + \big[\mathfrak{L}_{+} , \mathfrak{L}_{-}\big]\,.
\end{align}
%
%
The equations of motion (\ref{eom-v1}) and (\ref{eom-v2}) are rewritten as
\begin{align}
\begin{split}
    0=&\,
    \delta \hat{v}_{+}\varphi(z)\pa_{\bar{z}}\big[\cL_{-} + 2\hat{v}_{-}\big]
    -\!\!
    \sum_{\hat{z}\in\{\pm1\}}2\pi i\delta(z-\hat{z})
    \Re\lim_{z\to\hat{z}}
    \bigg(32\varphi^2(z) \delta \hat{v}_{+} \hat{v}_{+}\tr(\overline{\hat{v}_-}\hat{v}_-)E'(\xi) \bigg)
    \,,\\
    0=&\,
    \delta \hat{v}_{-}\varphi(z)\pa_{\bar{z}}\big[ \cL_+ + 2\hat{v}_{+}\big]
    -\!\!
    \sum_{\hat{z}\in\{\pm1\}}2\pi i\delta(z-\hat{z})
    \Re\lim_{z\to\hat{z}}
    \bigg(32\varphi^2(z) \delta \hat{v}_{-} \hat{v}_{-}\tr(\overline{\hat{v}_+}\hat{v}_+)E'(\xi) \bigg)
    \,.
\end{split}
\label{v-eom}
\end{align}

\subsection{Reduction to the 2D action}
We now perform the reduction of the action (\ref{S-tot}) to a two-dimensional integrable field theory by partially substituting the solution of the equations of motion.
We first solve the bulk equations of motion (\ref{Lax-flat-2}) and (\ref{Lax-flat-3}) under the boundary conditions (\ref{Dirichlet}).
In order to perform a reduction to a 2-dimensional theory, we shall focus upon the following solution for which each term in (\ref{Lax-flat-2}) and (\ref{Lax-flat-3}) vanishes:\footnote{It may be possible to consider other solutions to derive a 2-dimensional theory. But we will not pursue that direction here.}
\begin{align}
    0= \varphi(z)\pa_{\bar{z}}\cL_{\pm}\,,\qquad
    0= \varphi(z)\pa_{\bar{z}}\hat{v}_{\pm}
    \label{eom-meromorphic}
\end{align}
${}^{\forall}z\in\mathbb{C}P^1$\,.
Noting the boundary condition (\ref{Dirichlet}), so that $\hat{v}_{\pm}|_{z=0,\infty}=\hat{g}^{-1}\cv_{\pm}\hat{g}|_{z=0,\infty}=0$\,, a general form of the solution for $v_{\pm}$ is given by
\begin{align}
    \hat{v}_{\pm}(z,\sigma^{\pm})=\frac{z}{1-z^2}v_{\pm}(\sigma^{\pm})\,.
    \label{v-solution}
\end{align}
Note here that with the solution (\ref{v-solution}), $\xi$ in (\ref{xi-def}) is not finite because $\varphi(z)^2\tr(\overline{\hat{v}_+}\hat{v}_+)\tr(\overline{\hat{v}_-}\hat{v}_-)$ scales as $O((1-z^2)^{-2})$ around $z=\pm$\,.
Since the action (\ref{S-pot}) includes $\xi$ as the argument of the potential, we need to regularize the solution to define a finite action.
We here employ the following regularization
\begin{align}\begin{split}
    \hat{v}_{+}(z,\sigma^{\pm};\alpha)=&\,
    \frac{1}{2}
    \left(\frac{1}{1-z} - \frac{1}{1+z}\big(1-e^{-|z+1|^2/\alpha}\big)\right)v_{+}(\sigma^{\pm})\,,
    \qquad z\sim-1\,,\\
    \hat{v}_{-}(z,\sigma^{\pm};\alpha)=&\,
    \frac{1}{2}
    \left(\frac{1}{1-z}\big(1-e^{-|z-1|^2/\alpha}\big) - \frac{1}{1+z}\right)v_{-}(\sigma^{\pm})\,,
    \qquad z\sim+1\,,
    \label{v-regularized}
\end{split}\end{align}
where $\alpha$ is real and positive.
Indeed, we can see the integral $\int dz\wedge d\bar{z}\,\delta(z\pm 1)E(\xi)$ is then finite
since the limits are taken as
\begin{align}\begin{split}
    \lim_{\alpha\to0}\lim_{z\to +1}\xi=&\,\lim_{z\to+1}\big(16\varphi^2(z)\tr(\overline{\hat{v}_+}\hat{v}_+)\tr(\overline{\hat{v}_-}\hat{v}_-)\big)
    \\
    =&\,
    \lim_{\alpha\to0}\lim_{z\to +1}\bigg[
    \frac{(1-z^2)^2}{z^4}
    \left(\frac{1}{1-z} - \frac{1}{1+z}\big(1-e^{-|z+1|^2/\alpha}\big)\right)^2
    \left(- \frac{1}{1+z}+O(|z-1|)\right)^2
    \\
    &\hspace{60pt}
    \tr(\overline{v_+}v_+)\tr(\overline{v_-}v_-)
    \bigg]
    \\
    =&\,\tr(v_+v_+)\tr(v_-v_-) = \nu\,,
    \\
    \lim_{\alpha\to0}\lim_{z\to -1}\xi=&\,\tr(v_+v_+)\tr(v_-v_-) = \nu\,.
\end{split}\label{xi-limit}
\end{align}
We can also confirm that the regularized solution (\ref{v-regularized}) reduces to the original one (\ref{v-solution}) in the $\alpha\to 0$ limit.

\medskip

We now need to carefully consider the order of limits in the regularization process.
Since the regularization (\ref{v-regularized}) is introduced to prevent the divergence in the integral, we have to take $\alpha\to 0$ after performing the integration in the action.
The proper sequence of steps is thus as follows:
first, take $\lim_{z\to\hat{z}}\;(\hat{z}\in\{\pm 1\})$, integrate over $\cM\times C$\,, and then remove the regularization by taking the limit $\alpha\to0$\,.
Note that the equations of motion (\ref{eom-meromorphic}) do not exactly hold for finite $\alpha$\,, but they become satisfied in the $\alpha\to0$ limit.

\medskip

The Lax form $\cL_{\pm}$\,, on the other hand, can have simple pole at $z=\pm1$\,, and thus the general form with this pole structure is given by
\begin{align}
    \cL_{\pm}= \frac{V_{\pm}}{1\pm z}+ U_{\pm}\,,
\end{align}
with some $U_{\pm},V_{\pm}:\cM\to\mathfrak{g}$\,.
Here, $U_{\pm}$ and $V_{\pm}$ take value in the real Lie algebra $\mathfrak{g}$ due to the reality condition $\tau\cL=\mu_{\rm t}^{*}\cL$\,.
Recalling the definition of $\cL$ (\ref{Lax-form-def}), the boundary condition (\ref{Dirichlet}) reads
\begin{align}
    U_{\pm}= \hat{g}^{-1}\pa_{\pm}\hat{g}\big|_{z=\infty}\,,\qquad
    V_{\pm}+U_{\pm}=  \hat{g}^{-1}\pa_{\pm}\hat{g}\big|_{z=0}\,.
\end{align}
Using the 2-dimensional gauge invariance (\ref{2d-gauge}), we can always set the value $\hat{g}|_{z=\infty}=1$\,, and thus $\hat{g}^{-1}\pa_{\pm}\hat{g}|_{z=\infty}=0$\,.
After all, defining 
\begin{align}
    g:=\hat{g}|_{z=0}\,,
\end{align}
we can obtain a solution of the boundary condition as
\begin{align}
    \cL =&\, \frac{g^{-1}\pa_{+}g}{1+z}d\sigma^{+} + \frac{g^{-1}\pa_{-}g}{1-z}d\sigma^{-}\,,
    \label{L-solution}
    \\
    \mathfrak{L}=&\,
    \left(\frac{g^{-1}\pa_{+}g}{1+z} + \frac{2z}{1-z^2}v_+\right)d\sigma^{+} 
    + \left(\frac{g^{-1}\pa_{-}g}{1-z} + \frac{2z}{1-z^2}v_-\right)d\sigma^{-}\,.
\end{align}

\medskip

Before substituting the solution (\ref{v-solution}) and (\ref{L-solution}) into the action, we express the equations of motion (\ref{v-eom}) in terms of $\cL$ and $\hat{v}$\,.
A careful treatment of the variations $\delta \mathsf{v}_{\pm}$ is required since the configurations of $\hat{v}_+$ and $\hat{v}_-$ allow singularities of $(z-1)^{-1}$ and $(z+1)^{-1}$, respectively, for consistency with (\ref{v-regularized}).
We also constrain the variations as $\delta \mathsf{v}_{+}=0$ at $z=-1$, and $\delta \mathsf{v}_{-}=0$ at $z=+1$  
as a boundary condition.
We then introduce scaled variations 
$\delta \mathsf{v}_{\pm}:= 
(1\pm z)^{-1}\delta \hat{v}_{\pm}$ to be consistent with (\ref{v-regularized}),
and the expressions (\ref{v-eom}) read
\begin{align}
    0=&\,
    \delta \mathsf{v}_{+}\frac{(1+z)^2}{z^2}\pa_{\bar{z}}\big[ \cL_- + 2\hat{v}_{-}\big]
    -\!\!
    \sum_{\hat{z}\in\{\pm1\}}2\pi i\delta(z-\hat{z})
    \Re\lim_{z\to\hat{z}}
    \bigg(32\varphi^2(z)\frac{1+z}{1-z}\delta \mathsf{v}_{+}
    \hat{v}_{+}\tr(\overline{\hat{v}_-}\hat{v}_-)E'(\xi) \bigg)
    \no\\
    =&\,
    \delta \mathsf{v}_{+}\frac{(1+z)^2}{z^2}
    \bigg(2\pi i\delta(z-1)g^{-1}\pa_{-}g  - \frac{e^{-|z-1|^2/\alpha}}{\alpha}v_- \bigg)
    \no\\
    &
    -2\pi i\delta(z-1)\lim_{z\to+1}\bigg(
    32\varphi^2(z)\frac{1+z}{1-z}\delta\mathsf{v}_{+}
    \frac{z}{1-z^2}
    v_+
    \tr(\overline{v_{-}}v_{-})
    \frac{1}{4(1+z)^2} E'(\xi)\bigg)
    \no\\
    =&\,
    \delta \mathsf{v}_{+}\frac{(1+z)^2}{z^2}
    \bigg(2\pi i\delta(z-1)g^{-1}\pa_{-}g  - \frac{e^{-|z-1|^2/\alpha}}{\alpha}v_- \bigg)
    -16\pi i\delta(z-1)\delta\mathsf{v}_{+}v_+
    \tr(v_{-}^2)E'(\nu)
    \no\\
    \to&\,
    8\pi i \delta(z-1)\delta \mathsf{v}_{+}
    \big[g^{-1}\pa_{-}g + v_{-} - 2v_{+}\tr(v_-^2) E'(\nu) \big]\,,
    \label{eom-vr1}
\end{align}
where we take $\alpha\to 0$ limit in the last line.
We utilized the fact that $v_{\pm}$ and $\delta \mathsf{v}_{\pm}$ take value in the real Lie algebra $\mathfrak{g}$ at $z\in\mathbb{R}$ because of the reality condition (\ref{reality}).
Note here that as $z$ approaches the defects at $z=\pm1$, the variable $\xi$ tends to $\nu$ because of (\ref{xi-limit}).
In the second terms of the second equality in (\ref{eom-vr1}), we have replaced $\tr(\overline{\hat{v}_-}\hat{v}_{-})$ by $\frac{1}{4(1+z)^2}\tr(\overline{v_-}v_-)$. This calculation is justified since around $z=+1$ we have
\begin{align}
    \hat{v}_-=&\, -\frac{1}{2(1+z)}v_{-} + O(|z-1|)
\end{align}
due to (\ref{v-regularized}), and the contribution from the second term vanishes in (\ref{eom-vr1}).
On the other hand, $\hat{v}_+$ is replaced by $\frac{z}{1-z^2}v_+$ the regularization for $\hat{v}_+$ has no effects around $z=+1$ after taking the $\alpha\to 0$ limit.
In exactly the same way, the equation of motion for the other component is obtained as follows:
\begin{align}
    0=&\,
    \delta \mathsf{v}_{-}\frac{(1-z)^2}{z^2}\pa_{\bar{z}}\big[ \cL_+ + 2\hat{v}_{+}\big]
    -\!\!
    \sum_{\hat{z}\in\{\pm1\}}2\pi i\delta(z-\hat{z})
    \Re\lim_{z\to\hat{z}}
    \bigg(32\varphi^2(z)\frac{1-z}{1+z}\delta \mathsf{v}_{-}
    \hat{v}_{-}\tr(\overline{\hat{v}_+}\hat{v}_+)E'(\xi) \bigg)
    \no\\
    \to&\,
    8\pi i \delta(z+1)\delta \mathsf{v}_{-}
    \big[-g^{-1}\pa_{+}g + v_{+} - 2v_{-}\tr(v_+^2) E'(\nu) \big]\,,
    \label{eom-vr2}
\end{align}

\medskip

We now substitute the solutions (\ref{v-solution}) and (\ref{L-solution}) to obtain the effective 2-dimensional action.
The result is obtained as~\cite{Delduc:2019whp,Benini:2020skc}
\begin{align}
    S_{\rm 4dCS}=&\,
    \frac{1}{2}\int_{\cM}\tr(\operatorname{res}_{0}(\varphi(z)\cL)\wedge g^{-1}dg)
    \no\\
    =&\,
    -\int_{\cM}d\sigma^{+}\wedge d\sigma^{-}
    \tr(g^{-1}\pa_{+}gg^{-1}\pa_{-}g)
    \\
    S_{\rm int}=&\,
    \frac{i}{\pi}\int_{\cM\times C}
    \omega\wedge d\bar{z}\wedge\tr(\pa_{\bar{z}}\cL\wedge \hat{v})
    -\frac{i}{\pi}\int_{\cM\times C}
    \omega\wedge d\bar{z}\wedge\tr(\hat{v}\wedge\pa_{\bar{z}}\hat{v})
    \no\\
    =&\,
    -2\int_{\cM}d\sigma^{+}\wedge d\sigma^{-}
    \big(\tr(g^{-1}\pa_{+}g\,v_-)-\tr(g^{-1}\pa_{-}g\,v_+)\big)
    +2\int_{\cM}d\sigma^{+}\wedge d\sigma^{-}
    \tr(v_+v_-)
    \label{2d-int}
    \\
    \tilde{S}_{\rm pot}=&\,
    -2\int_{\cM} d\sigma^+\wedge d\sigma^-\:E(\nu)\,,\qquad
    \nu:=\tr(v_+v_+)\tr(v_-v_-)\,,
\end{align}
To derive the second line in (\ref{2d-int}), we need to consider the regularization (\ref{v-regularized}) and utilize the relations such as $(1-e^{-|z\pm1|^2/\alpha})\delta(z\pm1)=0$\,.
The effective potential term $\tilde{S}_{\rm pot}$ can be obtained so that the precise variation $\delta S_{\rm pot}[\cv_{\pm}]\propto\delta\tilde{S}_{\rm pot}[v_{\pm}]$ is reproduced with the constraint $\delta{\cv}_{\pm}=0$ at $z=\mp1$\,, respectively.
We can see that this two-dimensional effective action is exactly same as the AFSM action up to multiplication, i.e.,
\begin{align}
    -2 S_{\rm 2d}[g,v_{\pm}] \simeq S_{\rm 4dCS}[A]+ S_{\rm int}[A,\cv_{\pm}] + \tilde{S}_{\rm pot}[v_{\pm}]\,,
\end{align}
where $\simeq$ here denotes the equality that holds after substituting the solution (\ref{v-regularized}) and (\ref{L-solution}).

\subsection{Auxiliary field sigma model with the Wess-Zumino term}

Next, let us consider to include the Wess-Zumino term in the previous analysis. In the case of the principal chiral model, it is known that the twist function $\varphi(z)$ should be deformed as \cite{Delduc:2019whp}
\begin{align}
    \varphi(z)=\frac{1-z^2}{(z-k)^2}\,,
    \label{twist-mod}
\end{align}
with a real number $k$\,. Hence, we will follow the same strategy for the AFSM in the following. 

\medskip

The argument of the potential term $\xi$ is also modified as 
\begin{align}
    \xi := 
    16\varphi^2(z)\tr(\cv_+\cv_+)\tr(\cv_-\cv_-)
    =
    16\frac{(1-z^2)^2}{(z-k)^4}\tr(\cv_+\cv_+)\tr(\cv_-\cv_-)\,.
\end{align}
By considering these changes, the location of the Dirichlet boundary condition is shifted as 
\begin{align}\begin{split}
    A_{\pm}\big|_{z=k,\infty}=&\,0\,,\qquad 
    \\
    \cv_{\pm}\big|_{z=k,\infty}=&\,0\,.
    \label{Dirichlet-k}
\end{split}\end{align}
Then the gauge transformations should satisfy $u|_{z=k,\infty}=1$ instead of (\ref{u-condition}).
Under these boundary conditions,
the configuration 
\begin{align}\begin{split}
    \hat{v}_{+}(z,\sigma^{\pm};\alpha)=&\,
    \frac{1}{2}
    \left(\frac{1-k}{1-z} - \frac{1+k}{1+z}\big(1-e^{-|z+1|^2/\alpha}\big)\right)v_{+}(\sigma^{\pm})\,,
    \qquad z\sim-1\,,\\
    \hat{v}_{-}(z,\sigma^{\pm};\alpha)=&\,
    \frac{1}{2}
    \left(\frac{1-k}{1-z}\big(1-e^{-|z-1|^2/\alpha}\big) - \frac{1+k}{1+z}\right)v_{-}(\sigma^{\pm})\,,
    \qquad z\sim+1\,,
    \label{v-regularized-mod}
\end{split}\end{align}
gives the solution to (\ref{eom-meromorphic}) in the $\alpha\to 0$ limit although (\ref{v-regularized-mod}) itself is not an exact solution.

The Lax form is given by 
\begin{align}
    \cL =&\, (1+k)\frac{g^{-1}\pa_{+}g}{1+z}d\sigma^{+} + (1-k)\frac{g^{-1}\pa_{-}g}{1-z}d\sigma^{-}\,. 
    \label{L-solution-mod}
\end{align}
Upon obtaining these solutions, we can write down the equations of motion with respect to $\delta \cv$ as 
\begin{align}
    0=&\,
    \delta \hat{v}_{+}\varphi(z)\pa_{\bar{z}}\big[\cL_{-} + 2\hat{v}_{-}\big]
    -\!\!
    \sum_{\hat{z}\in\{\pm1\}}2\pi i\delta(z-\hat{z})
    \Re\lim_{z\to\hat{z}}
    \bigg(32\varphi^2(z) \delta \hat{v}_{+}\hat{v}_{+}\tr(\overline{\hat{v}_-}\hat{v}_-)E'(\xi) \bigg) 
    \no\\
    =&\,
     \frac{8\pi i}{1-k} \delta(z-1)\delta \mathsf{v}_{+}
    \big[g^{-1}\pa_{-}g + v_{-} - 2v_{+}\tr(v_-^2) E'(\nu) \big]\,,
    \label{WZ-auxiliary-1}
    \\
    0=&\,
    \delta \hat{v}_{-}\varphi(z)\pa_{\bar{z}}\big[ \cL_+ + 2\hat{v}_{+}\big]
    -\!\!
    \sum_{\hat{z}\in\{\pm1\}}2\pi i\delta(z-\hat{z})
    \Re\lim_{z\to\hat{z}}
    \bigg(32\varphi^2(z) \delta \hat{v}_{-} \hat{v}_{-}\tr(\overline{\hat{v}_+}\hat{v}_+)E'(\xi) \bigg)
    \no\\
    =&\,
    \frac{8\pi i}{1+k} \delta(z+1)\delta \mathsf{v}_{-}
    \big[-g^{-1}\pa_{+}g + v_{+} - 2v_{-}\tr(v_+^2) E'(\nu) \big]\,,
    \label{WZ-auxiliary-2}
\end{align}
corresponding to (\ref{eom-vr1}) and (\ref{eom-vr2}).
The reduction to a 2D action can also be performed as
\begin{align}
    S_{\rm 4dCS}\simeq&\,
    \frac{1}{2}\int_{\cM}\tr(\operatorname{res}_{k}(\varphi(z)\cL)\wedge g^{-1}dg)
    -\frac{1}{2}\operatorname{res}_{k}\big(\varphi(z)\big) I_{\rm WZ}
    \no\\
    =&\,-\int_{\cM} d\sigma^+\wedge d\sigma^-
    \tr(g^{-1}\pa_{+}gg^{-1}\pa_{-}g)
    + k I_{\rm WZ}\,,
    \\
    S_{\rm int}=&\,
    \frac{i}{\pi}\int_{\cM\times C}
    \omega\wedge d\bar{z}\wedge\tr(\pa_{\bar{z}}\cL\wedge \hat{v})
    -\frac{i}{\pi}\int_{\cM\times C}
    \omega\wedge d\bar{z}\wedge\tr(\hat{v}\wedge\pa_{\bar{z}}\hat{v})
    \no\\
    \simeq&\,
    -2\int_{\cM}d\sigma^{+}\wedge d\sigma^{-}
    \big(\tr(g^{-1}\pa_{+}g\,v_-)-\tr(g^{-1}\pa_{-}g\,v_+)\big) \no\\
    &
    +2\int_{\cM}d\sigma^{+}\wedge d\sigma^{-}
    \tr(v_+v_-)\,,
    \\
    \tilde{S}_{\rm pot}=&\,
    -2\int_{\cM} d\sigma^+\wedge d\sigma^-\:E(\nu)\,,\qquad
    \nu:=\tr(v_+v_+)\tr(v_-v_-)\,,
    \\
    S_{{\rm ASFM},k}[g,v_{\pm}]:=&\,
     \int_{\cM}d\sigma^+\wedge d\sigma^- 
    \bigg(
    \frac{1}{2}\tr(j_+j_-) - \tr(v_+v_-) + \tr(j_+v_- - j_-v_+) + E(\nu)
    \bigg) \no \\ 
   & -
    \frac{k}{2}I_{\rm WZ}
    \label{2d-action-mod}
    \\
    \simeq&\,
    -\frac{1}{2}\big(S_{\rm 4dCS} + S_{\rm int} + \tilde{S}_{\rm pot} \big)\,,
    \no
\end{align}
where the Wess-Zumino term $I_{\rm WZ}$ is defined as 
\begin{align}
    I_{\rm WZ}:=
    -\frac{1}{3}\int_{\cM\times [0,\varepsilon]}\tr(\hat{g}^{-1}d\hat{g}\wedge \hat{g}^{-1}d\hat{g} \wedge \hat{g}^{-1}d\hat{g})\,,
    \qquad 
    \varepsilon>0
\end{align}
with $\hat{g}(\sigma^{\pm};0)=g(\sigma^{\pm})$ and $\hat{g}(\sigma^{\pm};\varepsilon)=\bm{1}$\,.

\medskip 

As in the case of the original AFSM (\ref{action-AFSM}), parts of the equations of motion for the action $S_{\mathrm{ASFM},k}$ are indeed encoded by the flatness condition of the modified Lax pair
\begin{align}
    \mathfrak{L}=&\,
    \left((1+k)\frac{g^{-1}\pa_{+}g}{1+z} + \frac{2(z-k)}{1-z^2}v_+\right)d\sigma^{+} 
    + \left((1-k)\frac{g^{-1}\pa_{-}g}{1-z} + \frac{2(z-k)}{1-z^2}v_-\right)d\sigma^{-}
    \\
    =&\,
    \frac{ j_+ - k\mathfrak{J}_+ + z(\mathfrak{J}_{+}-kj_+)}{1-z^2}d\sigma^+
    +
    \frac{ j_- + k\mathfrak{J}_- - z(\mathfrak{J}_{-}+kj_-)}{1-z^2}d\sigma^-
\end{align}
with the modified current $\mathfrak{J}_{\pm}=- \big(j_{\pm} \mp 2 v_{\pm}\big)$\,. 
By using the auxiliary relations (\ref{jj-relation}) and the off-shell flatness condition (\ref{off-shell-flatness}), the field strength for $\mathfrak{L}_{\pm}$ is evaluated as follows: 
\begin{align}
     \pa_{+}&\mathfrak{L}_{-} - \pa_{-}\mathfrak{L}_{+} + [\mathfrak{L}_{+},\mathfrak{L}_-]
    \no\\
    =&\,
    \frac{\pa_{+}(j_{-}+k\mathfrak{J}_-)-\pa_{-}(j_{+}-k\mathfrak{J}_+)
    -z \big(\pa_{+}(\mathfrak{J}_{-}+kj_-) + \pa_{-}(\mathfrak{J}_{+}-kj_+)
    \big)}{1-z^2}
    \no\\
    &+
    \frac{[j_{+}-k\mathfrak{J}_+,j_{-}+k\mathfrak{J}_-] 
    - z[j_{+}-k\mathfrak{J}_+,\mathfrak{J}_{-}+kj_-]}{(1-z^2)^2} 
 \no \\     
    &+ \frac{z[\mathfrak{J}_{+}-kj_+,j_{-}+k\mathfrak{J}_-]
    - z^2 [\mathfrak{J}_{+}-kj_+,\mathfrak{J}_{-}+kj_-]}
    {(1-z^2)^2}
    \no\\
    \,\dot{=}\,&\,
    \frac{\pa_{+}(j_{-}+k\mathfrak{J}_-)-\pa_{-}(j_{+}-k\mathfrak{J}_+)
    -z \big(\pa_{+}(\mathfrak{J}_{-}+kj_-) + \pa_{-}(\mathfrak{J}_{+}-kj_+)
    \big)}{1-z^2}
    +\frac{(1-k^2)[j_+,j_-] 
    }{1-z^2}
    \no\\
    =&\,
    \frac{k-z}{1-z^2}
    \big( \pa_{+}(\mathfrak{J}_{-}+kj_-) + \pa_{-}(\mathfrak{J}_{+}-kj_+) \big)
    \,.
    \label{WZ-flatness}
\end{align}
From the classical action $S_{\mathrm{AFSM},k}$\,, the equations of motion can be derived as 
\begin{align}
    0\:\dot{=}\:&\,\pa_{+}(\mathfrak{J}_{-}+kj_-) + \pa_{-}(\mathfrak{J}_{+}-kj_+) \,.
\end{align}
Thus, the flatness condition for the Lax pair $\mathfrak{L}_{\pm}$ is indeed satisfied.

\section{Conclusion and Discussion}\label{sec:conclusion}
We have discussed the AFSM from the viewpoint of the 4D CS theory by introducing the auxiliary one-form $\cv=\cv_+ d\sigma^+ +\cv_- d\sigma^-$\,.
The total action (\ref{S-tot}) includes the interaction terms and the potential term, and respects the gauge symmetry under an appropriate boundary condition.
We then derived the equations of motion for the model, which allows us to perform the reduction to the 2D action.
As a result, we have correctly obtained the AFSM action and the Lax pair that leads to the equations of motion for the AFSM.

\medskip

We further derived the AFSM with the Wess-Zumino term by deforming the twist function as in (\ref{twist-mod}) with a real constant $k$.
Since the pole structure of the twist function is deformed, the form of the equations of motion should be modified as well.
The explicit forms of $\hat{v}$ and $\cL$ are given in (\ref{v-regularized-mod}) and (\ref{L-solution-mod}), respectively. 
Remarkably, the resulting 2D action (\ref{2d-action-mod}) just contains the Wess-Zumino term in comparison to the original AFSM action. 
The equation of motion for the AFSM with the Wess-Zumino term 
can be reproduced from the flatness condition of the Lax pair (\ref{WZ-flatness}) and the auxiliary equations (\ref{WZ-auxiliary-1}) and (\ref{WZ-auxiliary-2})

\medskip

As an outlook, it would be interesting to consider integrable deformations involving the stress tensor explicitly from the 4D CS perspective.
In \cite{Ferko:2024ali}, significant classes of deformations such as the original $T\overline{T}$-deformation and the root-$T\overline{T}$-deformation are obtained by eliminating the auxiliary fields $v_{\pm}$ by substituting the solution for a classical flow equation. Along this line, one may anticipate can that a similar flow equation should exist at the 4D CS level. It would be intriguing if we could reveal the 4D CS origin of the flow equation. 

\medskip

It is also possible to perform other integrable deformations of the 4D CS action (\ref{2d-action-mod}). By following the work \cite{Delduc:2019whp}, Yang-Baxter deformations and their cousins are realized by deforming the pole structure of the twist function and changing the boundary conditions so as to involve the $R$ operators. Thus we anticipate that Yang-Baxter deformations of the AFSM can be discussed by the similar technique. We will report the results in another place \cite{FY}.

\subsection*{Acknowledgments}

The work of O.\,F.\ was supported by RIKEN Special Postdoctoral Researchers Program. 
The work of K.Y. was supported by MEXT KAKENHI Grant-in-Aid for Transformative Research Areas A “Machine Learning Physics” No.\,22H05115, and JSPS Grant-in-Aid for Scientific Research (B) No.\,22H01217.



\bibliographystyle{utphys}
\bibliography{auxiliary}


\end{document}